%% file: HST_LBQS0302.tex
\documentclass{aa}  
\usepackage{graphicx}
\usepackage{txfonts}
\usepackage{hyperref}
\usepackage{color}

\begin{document}

  \title{
  A meeting at $z\sim3$: Young massive galaxies and an AGN within 30\,kpc of the luminous QSO LBQS~0302$-$0019}
  \titlerunning{Young massive galaxies, and an AGN within 30\,kpc from the luminous QSO LBQS~0302$-$0019}
  \author{B.~Husemann\inst{1}
         \and
          G.~Worseck\inst{2}
         \and
         F.~Arrigoni Battaia\inst{3}
         \and 
         A.~A.~C.~Sander\inst{4}
         \and
         T.~Shanks\inst{5}
        }

  \institute{
    Max-Planck-Institut f\"ur Astronomie, K\"onigstuhl 17, D-69117 Heidelberg, Germany,  \email{husemann@mpia.de}
    \and
    Institut f\"ur Physik und Astronomie, Universit\"at Potsdam, Karl-Liebknecht-Str. 24/25, D-14476 Potsdam, Germany
    \and
    Max-Planck-Institut f\"ur Astrophysik, Karl-Schwarzschild-Str 1, D-85748 Garching bei M\"unchen, Germany
    \and
    Armagh Observatory and Planetarium, College Hill, Armagh BT61 9DG, N. Ireland
    \and
    Centre for Extragalactic Astronomy, Durham University, South Road, Durham, DH1 3LE, UK
   }

   \date{}

  \abstract
   {
   Contrary to expectations from scenarios of black hole growth driven by galaxy interactions and mergers, dual active galactic nuclei (AGN) with kiloparsec separations are rarely observed and are very difficult to identify, in particular at high redshifts (i.e. $z>2$).
   }
   {Focussing on the recently discovered dual AGN system LBQS~0302$-$0019 at $z=3.29$, we seek to identify further group members in its environment and to understand their formation history through deep high-angular-resolution imaging. }
   {We present deep \textit{Hubble Space Telescope (HST)} Wide-field Camera~3 near-infrared imaging of LBQS~0302$-$0019. In combination with ground-based VLT/HAWK-I imaging, we infer accurate sizes, colours, ages, and stellar masses of companion galaxies. }
   {We clearly detect four companion objects close to LBQS~0302$-$0019 that also have faint signatures in the ground-based images. We constrain light-weighted ages and masses for the two most prominent companions, Jil1 and Jil2, to  $t_{\star}=252_{-109}^{+222}$\,Myr with $\log(M_\star/[\mathrm{M}_\odot])= 11.2_{-0.1}^{+0.3}$ and $t_{\star}=19_{-14}^{+74}$\,Myr with $\log(M_\star/[\mathrm{M}_\odot])= 9.4_{-0.4}^{+0.9}$, respectively. The \textit{HST} data also show that the obscured AGN, previously identified by strong nebular \ion{He}{ii} emission, is associated with the young massive companion Jil2. Because very massive stars of the starburst cannot be solely responsible for the \ion{He}{ii} emission, we strengthen our initial conclusion that Jil2 has been hosting an AGN.}
   {
   If the young starburst of Jil2 had been accompanied by sustained black hole growth, Jil2 may have contributed \ion{He}{ii}-ionising flux to create the large \ion{He}{ii} Ly$\alpha$ proximity zone around LBQS~0302$-$0019. Hence, the duration of the current luminous AGN episode of LBQS~0302$-$0019 may have been overestimated.
   }


   \maketitle
%

\section{Introduction}
During the evolution of massive galaxies, their nuclei are expected to show recurrent phases of activity in order to grow their super-massive black holes (SMBHs) as observed from the cosmic accretion rate distribution  \citep[e.g.][]{Soltan:1982,Marconi:2004, Schulze:2015,Georgakakis:2017}.
While mergers of galaxies were initially thought to be the dominant mechanism triggering such activity \citep[e.g.][]{Sanders:1988a, Hopkins:2008a}, this has been strongly debated in recent years \citep[e.g.][]{Mechtley:2016,Villforth:2017,Weigel:2018,Marian:2019}. Nevertheless, to confirm the former scenario, a strong interest has been growing to identify so-called dual and binary active galactic nuclei (AGN), where two SMBHs with separations of 100\,kpc down to a few parsecs are active simultaneously \citep[see][for a recent review]{deRosa:2019}. The fraction of confirmed dual AGN is small even at low redshifts, and only a few dozen systems are known to date \citep[e.g.][]{Komossa:2003,Guainazzi:2005,Bianchi:2008,Piconcelli:2010,Koss:2011,Fu:2012,Woo:2014,Gross:2019,Husemann:2020,Silverman:2020}.

Detecting dual AGN at high redshifts ($z>2$) is increasingly difficult due to the limitations in sensitivity, spatial resolution, and area covered by X-ray surveys, optical, near-infrared (NIR) spectroscopic surveys, and radio surveys. Nevertheless, massive galaxy evolution at early times
mainly occurs in high-density environments, which is consistent with our hierarchical galaxy evolution picture \citep[e.g.][]{Venemans:2007, Wylezalek:2013, Hatch:2014, Stott:2020}. Unsurprisingly, galaxy overdensities and frequent companion galaxies have been detected around luminous
quasi-stellar objects (QSOs) at high redshifts \citep[e.g.][]{Decarli:2017,Trakhtenbrot:2017, Venemans:2020}. Still, the number of reported dual AGN systems at $2<z<7$ with $<$50\,kpc separation has been sparse so far, with only a few confirmed cases \citep{Djorgovski:2007, Hennawi:2015, ArrigoniBattaia:2018, Findlay:2018}.

The luminous QSO LBQS~0302$-$0019 at $z=3.29$ has been extensively studied to infer the properties of the intergalactic medium (IGM) along our line of sight \citep[e.g.][]{Hu:1995,Steidel:2003,Jakobsen:2003,Tummuangpak:2014,Schmidt:2017}. It is one of the rare UV-transparent luminous QSOs that allow the \ion{He}{ii} Ly$\alpha$ absorption of the IGM to be investigated in detail together with the proximity zone caused by the enhanced ionising photon flux around the QSO  \citep[e.g.][]{Jakobsen:1994,Syphers:2014}. A large proximity zone of 13.2\,Mpc is determined for LBQS~0302$-$0019 \citep{Worseck:2021}, which implies a long AGN phase of $>$11\,Myr so far.

Analyzing archival Multi-Unit Spectroscopic Explorer (MUSE) observations of LBQS~0302$-$0019, \citet[][hereafter Paper I]{Husemann:2018} reported the serendipitous discovery of an obscured AGN about 20\,kpc away from the QSO based on Ly$\alpha$, \ion{C}{iv} $\lambda\lambda 1548,1550$, \ion{He}{ii}$\lambda$1630, and [\ion{C}{iii}] $\lambda\lambda1907,1909$ UV emission-line diagnostics. In particular, the \ion{He}{ii} line luminosity $\log(L_{\ion{He}{ii}}/[\mathrm{erg\,s}^{-1}])=42.24\pm0.05$ 
was inconsistent with being induced by LBQS~0302$-$0019 given the compact, point-like spatial distribution of the emission and its corresponding low cross-section. The \ion{He}{ii} luminosity can be more easily explained by an AGN of around 1/500--1/1000 the luminosity of LBQS~0302$-$0019 (corresponding to $L_\mathrm{AGN}\sim10^{45}\,\mathrm{erg\,s}^{-1}$) if located within the compact \ion{He}{ii} region, where the major uncertainty is the distance of the recombining clouds to the embedded AGN. Radio observations could not identify radio emission associated with the obscured AGN \citep{Frey:2018}, dubbed Jil, but those observations are too shallow to detect an AGN at the estimated luminosity.  Follow-up ground-based $K_\mathrm{s}$-band imaging and NIR spectroscopy have been presented in \citet[][hereafter Paper II]{Husemann:2018c}, which successfully detected Jil's host galaxy with rest-frame optical line ratios consistent with an obscured AGN interpretation. While the NIR image obtained with the High Acuity Wide field K-band Imager (HAWK-I) at the Very Large Telescope (VLT) reveals signatures of distortion in the morphology, a clear interpretation is limited by the depth and resolution of the ground-based observations. Nevertheless, these observations implied massive host galaxies for both AGN of about $10^{11}M_\sun$, corresponding to a large overdensity at redshift $z=3.29$.

\begin{figure*}
 \includegraphics[width=\textwidth]{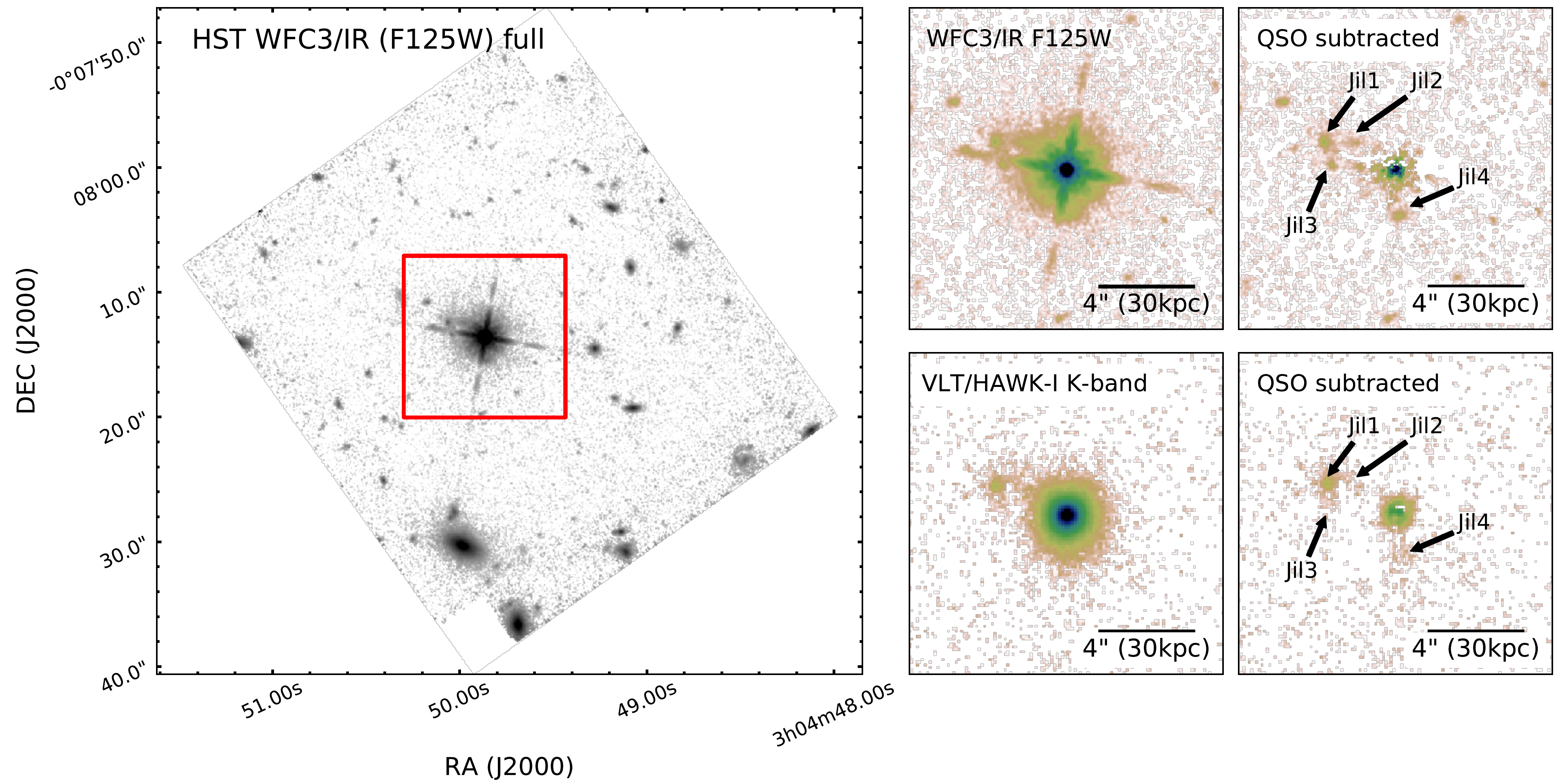}
 \caption{
 \textit{HST} WFC3/IR (F125W) image of the QSO LBQS~0302$-$0019 in comparison to ground-based adaptive-optics-assisted VLT/HAWK-I ($K_\mathrm{s}$) imaging. \textit{Left panel:} WFC3 image after combination of all frames taken in the IRSUB256 sub-aperture. North is up, and east is to the left. \textit{Right panels:} Zoomed-in view on the central $13\arcsec\times13\arcsec$ around the QSO in the \textit{HST} (top) and HAWK-I (bottom) images from Paper II before (left) and after (right) QSO subtraction. Four close neighbouring galaxies are consistently seen in both filters and are annotated as Jil1--4. }\label{fig:overview}
\end{figure*}

In this paper we combine our previous observations from MUSE and HAWK-I with a deep Hubble Space Telescope (HST) Wide-field Camera~3 (WFC3) NIR image of this system. Combining all the information allows us to provide new details on the morphology and the stellar populations of the galaxies as well as a more refined association of the rest-frame UV emission lines with the host galaxies. Throughout the paper we assume a concordance cosmology with $H_0=70\,\mathrm{km\,s}^{-1}\mathrm{Mpc}^{-1}$, $\Omega_m=0.3$, and $\Omega_\Lambda=0.7$.

\section{Deep HST WFC3/IR imaging of LBQS~0302$-$0019}
\subsection{Observations and data reduction}
LBQS~0302$-$0019 was observed on 2018 August 30 with the infrared channel of WFC3 aboard \textit{HST} under programme GO15480 (PI: Bernd Husemann). We used the F125W filter corresponding to the $J$ band for which \textit{HST} provides an angular resolution of $\sim$0\farcs136, which is significantly under-sampled at the native WFC3/IR detector plate scale of $\sim$0\farcs13\,pixel$^{-1}$. Additionally, the QSO is bright, with a magnitude of $m_K=15.4$\,mag (Vega), which demands a dedicated observing strategy to achieve a high contrast and increased spatial sampling. Hence, we divided the four-orbit observations into a series of 40 short (44\,s) and 32 long (223\,s) dithered exposures as follows.

Given the small angular size of the overall system of a few arcseconds, we used the IRSUB256 sub-aperture of the detector to significantly reduce the data rate and buffer space needed for the many exposures. A two-point dither pattern with an offset of 5\farcs2 was combined with a rectangular four-point dither pattern to optimally oversample the point-spread function (PSF). The SPARS10 readout sequence with seven sampling points (i.e. 44\,s exposure time) and the SPARS25 readout sequence with eleven sampling points (i.e. 223\,s exposure time) was used for the short and long exposures, respectively. At each dither position we obtained five and four individual exposures for the short and long exposure sequences, respectively. This observing sequence accumulated 1760\,s of short exposures and 7136\,s of long exposures; the short exposure sequence was taken first to minimise detector persistence effects.

The automatic \textit{HST} archive pipeline processing only combines the frames of an individual dither position and does not combine all long and short observations into one optimally combined frame with increased sampling. We therefore made use of the individually calibrated frames provided by the automatic processing and drizzled all the long observations and all the short observations into two combined frames with the DrizzlePak package \citep{Avila:2015}, adopting an output sampling of 0\farcs06\,pixel$^{-1}$ with a drop size of $0.8$. To correct the bright QSO core emission of LBQS~0302$-$0019 from persistence effects, we replaced the count rates of the central $8\times8$ pixels ($0\farcs48\times0\farcs48$) in the combined long exposure with those of the combined short exposure. The fully combined \textit{HST} image is shown in Fig.~\ref{fig:overview}.

\subsection{Estimation of the point-spread function}
One critical aspect for the \textit{HST} image analysis is an accurate estimation of the PSF during the \textit{HST} observations. Because reconstructed PSFs from the WFC3/IR PSF library produced unacceptable residuals, we searched for archival WFC3/IR exposures in the F125W filter, which were taken close to our observations, to derive a fully empirical PSF. A stellar field centred on WISE 0830$-$6323 was observed a couple of orbits before our observations (GO15468, PI: Jacqueline Faherty). The observations consist of four dithered exposures obtained with the SPARS25 readout sequence and 16 sampling points. We combined the fully calibrated individual frames using DrizzlePak, employing the same settings as for our LBQS~0302$-$0019 observations. We extracted the images of four isolated bright stars and six fainter stars close to the centre of the field of view with offsets ranging from 10\arcsec\ to 50\arcsec\ with respect to the QSO camera position. We averaged the images of the bright and faint stars and replaced count rates of the central 8$\times$8 pixels at the core of the averaged bright star image with that of the averaged faint star image to avoid the impact of persistence on the PSF shape. In order to account for potential systematic uncertainties in the empirical PSF, we also averaged only three of the four bright stars and five of the six faint stars; this led to
24 different PSF combinations, which we later used to determine the uncertainties of the derived parameters.

\begin{table*}
 \small
 \caption{Physical parameters of identified companion galaxies.}\label{tab:galaxy_parameters}
 \input{Table1.tex}
\end{table*}

\subsection{PSF subtraction and identification of companions}
Based on the empirically constructed PSF for our \textit{HST}/WFC3 observations, we subtracted the QSO contribution from the image by scaling the entire PSF using \texttt{galfit}  \citep{Peng:2002,Peng:2010}. Unlike the processing of our deep VLT/HAWK-I image, we were unable to properly model the 2D surface brightness distribution of the QSO host galaxy. This was likely caused by the combination of several effects: 1) the F125W filter probes bluer wavelengths, where the contrast between QSO and host emission is much higher, 2) the PSF of \textit{HST} has a lot more substructure exactly on the expected host galaxy scales, and 3) the time difference and positional differences between the empirical PSF observations and our observations can cause subtle differences in the exact PSF shape. In any case, a simple PSF subtraction of the QSO is sufficient to uncover the position and brightness of several companion galaxies around LBQS~0302$-$0019, which is the primary aim of the observations. 

With the improved angular resolution and sensitivity of our \textit{HST} imaging, we find that the source initially termed `Jil' is resolved into three distinct galaxies with different sizes and brightnesses, as highlighted in Fig.~\ref{fig:overview}. Those components are consistent with the additional diffuse extended emission in the VLT/HAWK-I $K_\mathrm{s}$-band image around the location of Jil, presented in Paper II, which is now designated as Jil1. The components are separated by about 1\arcsec, corresponding to roughly 8\,kpc at the redshift of the QSO.
Given the separation of the components, it is much more likely that we see three individual galaxies than one galaxy with bright knots. In addition to the close complex of three galaxies to the east of LBQS~0302$-$0019, we already found tentative evidence in the VLT/HAWK-I image for another galaxy companion to the south. The \textit{HST} image confirms a distinct source at the expected location, which we designate as Jil4. Hence, we can report the clear detection of four companion galaxies within a radius of 30 kpc around the luminous QSO, and we explore their physical parameters below.

\begin{figure*}
 \includegraphics[width=\textwidth]{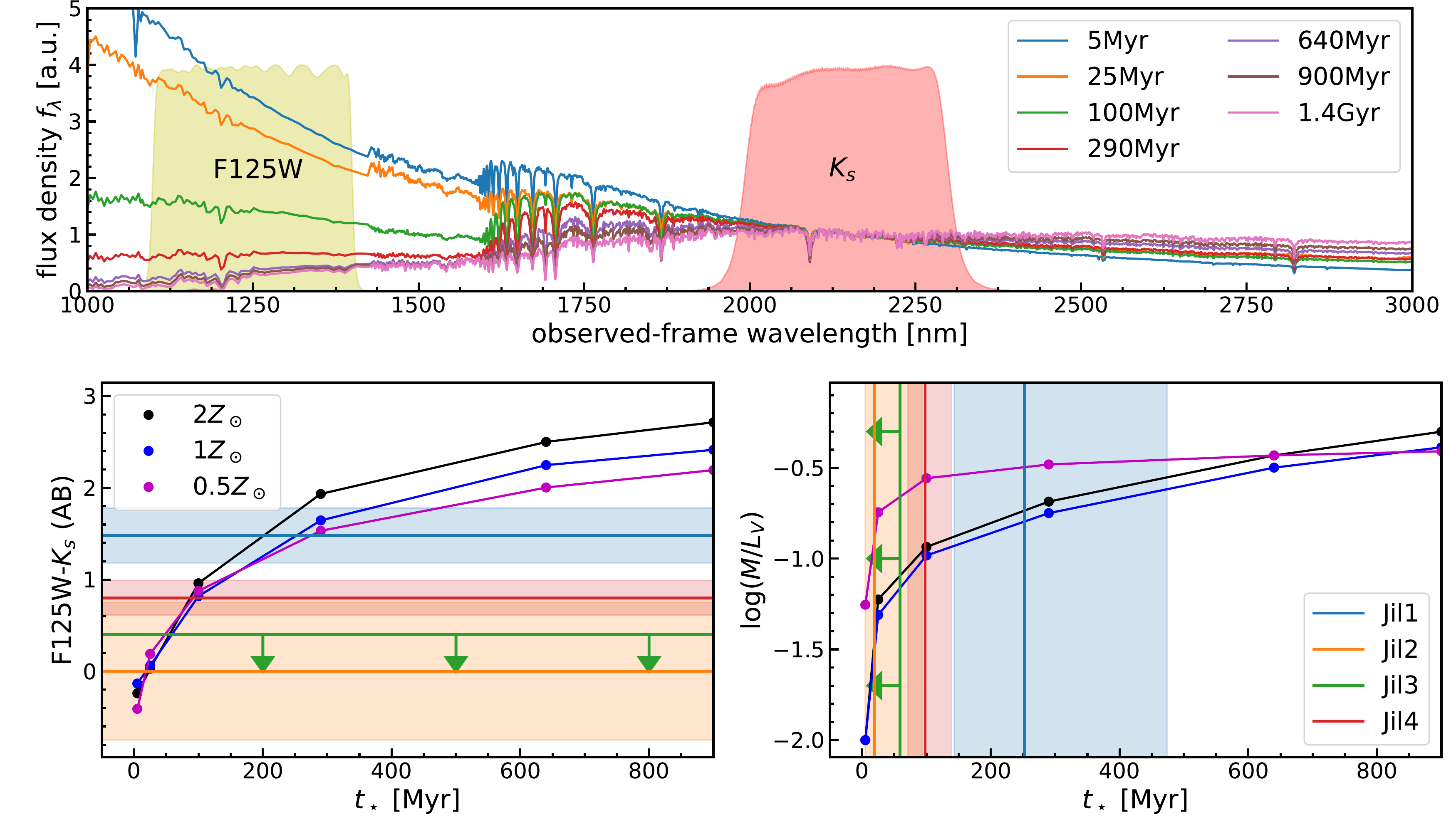}
 \caption{Estimation of stellar population age and stellar mass based on the observed F125W$-K_\mathrm{s}$ colour. \textit{Top panel:} Stellar population synthesis spectra from \citet{Bruzual:2003} redshifted to $z=3.29$ for various stellar ages in the wavelength range of interest. The spectra are normalised to a $K_\mathrm{s}$-band flux of unity. The VLT/HAWK-I $K_\mathrm{s}$ and WFC3/IR F125W filter curves are shown for comparison. \textit{Lower left panel:} Evolution of the observed F125W$-K_\mathrm{s}$ (roughly rest-frame $U-V$) colour with stellar ages up to 1\,Gyr and three metallicities. The measured colours and their uncertainty bands for Jil1, Jil2, and Jil4 are overplotted as horizontal stripes, while the upper limit for Jil3 is highlighted by arrows. \textit{Lower right panel:} Logarithmic mass-to-light ratio
 ($\log(M_\star[\mathrm{M}_\sun]/L_V[\mathrm{L}_\sun])$) as a function of stellar age. The confidence intervals in age for the galaxies are shown as vertical stripes, as determined from the stellar population synthesis spectra. The upper limit on the age for Jil3 is again highlighted by arrows.}\label{fig:colors}
\end{figure*}

\subsection{Morphological parameters and brightness}
We inferred the magnitude and morphological parameters for each component by adding single Sersi\'c components to the \texttt{galfit} model at the respective locations. Thereby, we inferred the brightness, effective radius, Sersi\'c index, axis ratio, and position angle for each component.  The measured parameters and their estimated uncertainties are listed in Table~\ref{tab:galaxy_parameters}. In order to determine the uncertainties, we performed Monte Carlo simulations and re-fitted the image 250 times with \texttt{galfit} after randomly drawing values from a Gaussian distribution centred on the initial value with the noise as standard deviation. One of the 24 empirical PSFs was used for each iteration to incorporate the systematic uncertainty of the PSF into the Monte Carlo simulations.

Afterwards, we also modelled the VLT/HAWK-I $K_\mathrm{s}$-band image with \texttt{galfit}, but keeping the relative positions, sizes, Sersi\'c indices, elongations, and position angles fixed to the best-fit values obtained from the \textit{HST} image. The only free parameters were the $K_\mathrm{s}$-band magnitudes of the components. We performed Monte Carlo simulations with \texttt{galfit} in the same manner as for the \textit{HST} image. Jil3 was not always clearly detected, so we determine a 99\% confidence lower limit. The $K_\mathrm{s}$ magnitudes and F125W-$K_\mathrm{s}$ colour terms are listed in Table~\ref{tab:galaxy_parameters}.

\section{Physical properties of the companion galaxies}
\subsection{Galaxy ages and stellar masses}
In order to determine more precise stellar masses than those predicted based on the single $K_\mathrm{s}$ band in Paper II, we compared the measured F125W$-K_\mathrm{s}$ colours with single stellar population (SSP) models from \citet{Bruzual:2003}. We show redshifted SSP spectra for various
stellar population ages in the upper panel of Fig.~\ref{fig:colors} and highlight the respective transmission curve of the WFC3/IR F125W and HAWK-I $K_\mathrm{s}$ filters for comparison. The observed F125W-$K_\mathrm{s}$ colour is very sensitive to the age of the SSP, as shown in the lower left panel of Fig.~\ref{fig:colors} for three different metallicities (half-solar, solar, and twice solar). From the observed colours, we inferred the age by linearly interpolating the age grid for solar metallicity and used the colour uncertainty and difference in metallicities to determine a $1\sigma$ confidence interval. The SSP ages correspond to a certain mass-to-light ($M_\star/L_V$) ratio for the rest-frame $V$ band, as shown in the lower left panel of Fig.~\ref{fig:colors}. For Jil3, we estimated corresponding upper limits for the age and $M_\star/L_V$ from the upper limit in colour and list all inferred parameters in Table~\ref{tab:galaxy_parameters}.

All galaxies appear young, with ages less than 300\,Myr, for the dominant light-weighted SSP. This is less than 20\% of the age of the Universe at $z=3.29$. Hence, the group of galaxies around LBQS~0302$-$0019 must have established recently and must have been undergoing intense star formation activity. The oldest galaxy with the reddest colours is the brightest component, Jil1. The high Sersi\'c index implies that this galaxy is compact and has likely already experienced a series of mergers, growing in mass,  in addition to continuous star formation. We used the mass-to-light ratios directly inferred from the colours to estimate a stellar mass for each component, which are listed in Table~\ref{tab:galaxy_parameters}. The stellar mass for Jil1 of $\log(M_\star/\mathrm{M}_\sun) = 11.2^{+0.3}_{-0.1}$ is fully consistent with our previous, less accurate estimate of $\log(M_\star/\mathrm{M}_\sun)=10.9\pm0.5$ from VLT/HAWK-I (Paper II). Notably, we discovered more than one companion with masses close to or greater than that of the Milky Way. This is consistent with our previous interpretation that LBQS~0302$-$0019 is associated with a group of galaxies within an exceptionally massive dark matter halo.

\subsection{Association with the Ly$\alpha$ and \ion{He}{ii} emission}
\begin{figure}
 \resizebox{\hsize}{!}{\includegraphics[]{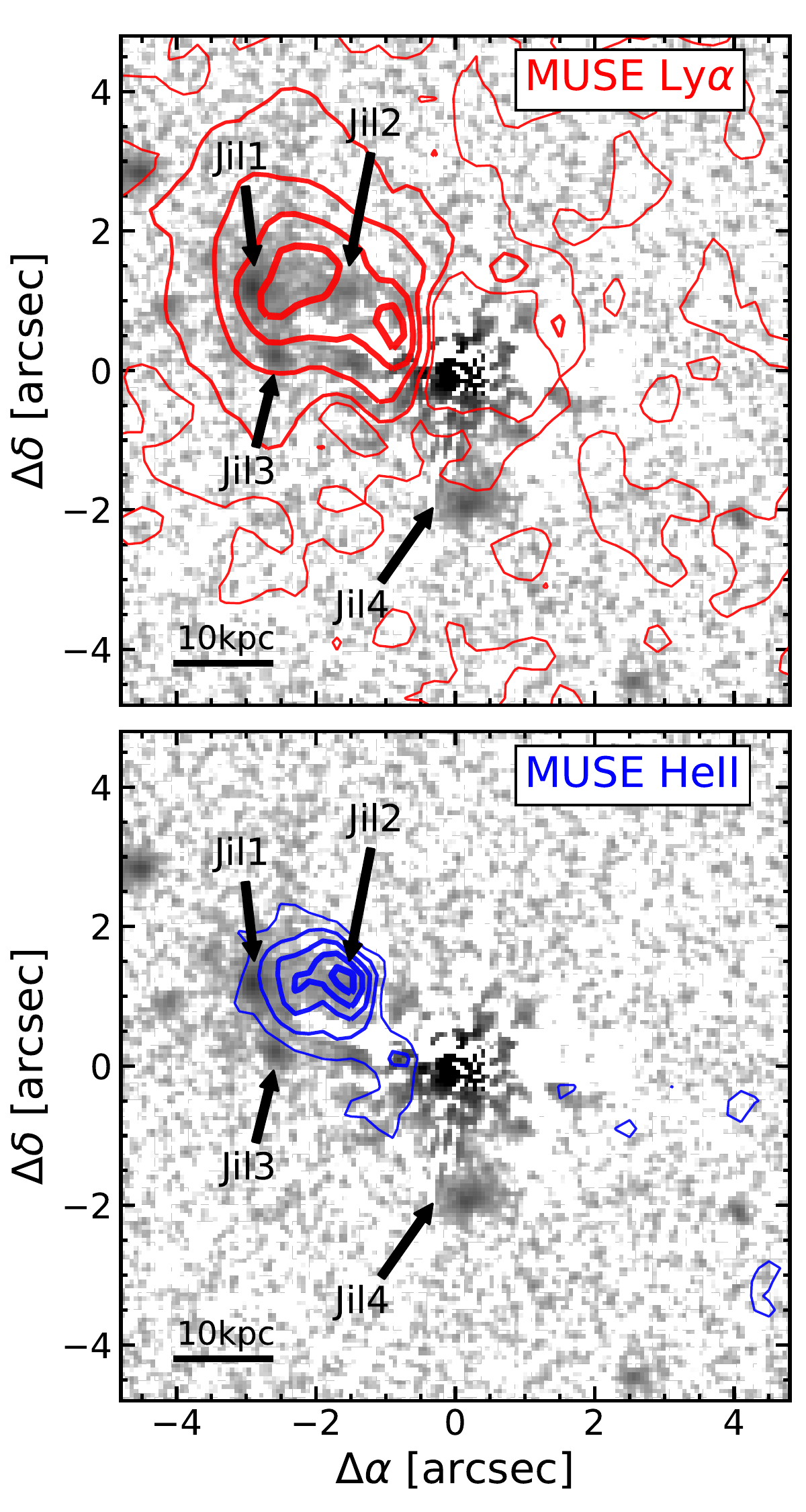}}
 \caption{Comparison of the Ly$\alpha$ and \ion{He}{ii} emission line distribution with the \textit{HST} continuum light. The QSO-subtracted WFC3 F125W image is shown in grey scale using a logarithmic scaling. The QSO-subtracted Ly$\alpha$ and \ion{He}{ii} centred narrow-band images obtained with MUSE are shown as red and blue contours in the upper and lower panel, respectively. Contour levels represent isophotes at $\Sigma_{\mathrm{Ly}\alpha}=[1, 2, 4, 6, 8]\times 10^{-17}\mathrm{erg\,s}^{-1}\mathrm{cm}^{-2}\mathrm{arcsec}^{-2}$ and $\Sigma_{\ion{He}{ii}}=[0.4, 0.6, 0.8, 1.0, 1.2]\times 10^{-17}\mathrm{erg\,s}^{-1}\mathrm{cm}^{-2}\mathrm{arcsec}^{-2}$, respectively. The Ly$\alpha$ extends slightly beyond the shown field at lower surface brightness levels. The MUSE maps were smoothed with a one-pixel-wide Gaussian to suppress noise, as presented in Paper I. The \textit{HST} and MUSE images are empirically registered based on the accurate position of the bright QSO.}\label{fig:HST_MUSE}
\end{figure}

The MUSE observations presented in Paper I were obtained as part of programme 094.A-0767(A) (PI: T. Shanks) and cover the wavelength range 4750\,\AA\ to 9300\,\AA\ over a $1\arcmin\times1\arcmin$ field of view at 0\farcs2 sampling. From the bright QSO continuum, a wavelength-dependent PSF was reconstructed using a median filter that effectively clipped narrow-line emission. This PSF was scaled in flux to match the brightest spaxel at the QSO position and subtracted from the data, which is a similar approach to what has been used in other works \citep[e.g.][]{Borisova:2016,Drake:2019}. More details of the MUSE data processing can be found in Paper I. Narrow-band images were extracted from the QSO-subtracted data in Paper I, which revealed an extended Ly$\alpha$ nebula around LBQS~0302$-$0019, as commonly observed around high-redshift QSOs \citep[e.g.][]{Heckman:1991b,Borisova:2016, ArrigoniBattaia:2019,Drake:2019,Farina:2019,OSullivan:2020}. In order to characterise the asymmetry of the nebula, we measured the asymmetry parameter, $\alpha=0.55$, as defined by \citet{ArrigoniBattaia:2019} and a displacement of the flux-weighted centroid of the Ly$\alpha$ nebula with respect to the QSO position of $d_\mathrm{QSO-Neb}=0\farcs76$ (5.7\,kpc) within the $2\sigma$ isophotal limiting area of $105.3\,\mathrm{arcsec}^2$. These values are fully consistent with the mean asymmetries and characteristics of Ly$\alpha$ nebulae studied by \citet{ArrigoniBattaia:2019}. Nevertheless, we had already concluded from the comparison of the HAWK-I and MUSE data in Paper II that the high surface brightness Ly$\alpha$ emission of the Ly$\alpha$ nebula is clearly associated with the neighbouring galaxy that is driving the asymmetry. The nearly point-like \ion{He}{ii} emission appeared slightly offset from the bright companion galaxy when comparing the HAWK-I and MUSE data. As highlighted in Fig.~\ref{fig:HST_MUSE}, the higher-resolution and higher-quality WFC3 image allows for a much more secure association of nebular line emission with the distinct companion galaxies.

The Ly$\alpha$ emission surface brightness peaks in between Jil1 and Jil2, which is likely caused by a superposition of individual Ly$\alpha$
nebulae, possibly together with emission from stripped material of the interacting galaxies. As noted by \citet{Wisotzki:2016}, all isolated star-forming galaxies at $z>3$ are surrounded by a Ly$\alpha$ halo with an effective radius about ten times larger than the one estimated from starlight. However, we noticed that no excess in Ly$\alpha$ emission is associated with Jil4 beyond the surface brightness of the diffuse Ly$\alpha$ symmetrically surrounding the QSO, as shown in Paper I. Hence, we cannot verify its physical association with the system despite the fact that the obtained colours imply sensible galaxy properties at the QSO redshift. Similarly, Jil3 is too close to Jil1 to separate the emission of both components, so Jil3 may simply be embedded within Jil1's extended Ly$\alpha$ nebula or might actually be at a different redshift. A chance projection could be an alternative explanation for  the extreme colours of Jil3, but a physical association cannot be ruled out with the current data.

With the \textit{HST} imaging we can conclusively associate all the narrow \ion{He}{ii} emission detected by MUSE with the companion Jil2 (Fig.~\ref{fig:HST_MUSE}) instead of the more massive galaxy Jil1. This is what we initially assumed from the HAWK-I image presented in Paper II despite the small but noticeable spatial offset. Given the high mass and very young age of Jil2, it is worth discussing whether stellar evolution alone might explain the associated nebular \ion{He}{ii} emission or if additional ionisation by an AGN is a necessity.

\subsection{The origin of \ion{He}{ii}\ $\lambda$1640 emission in Jil2}
Given the energy of 54\,eV required to fully ionise helium, the necessary hard radiation field is usually not produced by an evolved stellar population.
In addition to AGN, the radiation field of some types of hot, massive stars is sufficient to fully ionise helium. While this seems to naturally point towards classical Wolf-Rayet stars \citep[e.g.][]{Schaerer:1998}, their budget of \ion{He}{ii}-ionising flux can be effectively zero at high metallicity \citep{Sander:2020}. Instead, only Wolf-Rayet stars with weaker winds and stripped-envelope stars without strong winds manage to produce considerable amounts of \ion{He}{ii}-ionising flux. Consequently, the production of \ion{He}{ii}-ionising flux is fostered by lower metallicities as the lower available opacity shifts the onset of Wolf-Rayet-type winds to higher and higher masses \citep{Sander:2020}. As all these stars are very short lived; only galaxies undergoing a rare phase of intense star formation are capable of ionising significant amounts of helium. Examples are local low-metallicity dwarf galaxies \citep[e.g.][]{Kehrig:2015, Kehrig:2018, Senchyna:2020} or rare galaxies detected in deep spectroscopic galaxy surveys at intermediate redshift \citep[e.g.][]{Cassata:2013, Nanyakkara:2019, Saxena:2020}.

Wolf-Rayet stars typically show broad \ion{He}{ii} emission lines of $\sim$1000\,$\mathrm{km\,s}^{-1}$ full width at half maximum (FWHM) as part of their powerful stellar winds \citep[e.g.][]{Crowther:2007, Wofford:2014}. Hence, a prominent broad \ion{He}{ii} component is expected for a Wolf-Rayet galaxy in which such stars are a significant population \citep[e.g.][]{Brinchmann:2008, Miralles-Caballero:2016, Liang:2020}. The \ion{He}{ii} line detected with MUSE for Jil2 is narrow instead, with a FWHM of only $\sim300\mathrm{km\,s}^{-1}$, as characterised in Paper I, and therefore is incompatible with a traditional Wolf-Rayet galaxy interpretation. At low metallicity ($Z \leq 0.1 Z_\odot$), very massive stars with a WNh-type spectrum can efficiently produce narrow \ion{He}{ii} emission \citep{Graefener:2015}, but the reported line ratio of $\log(\ion{C}{iv}\lambda\lambda1548,1550/\ion{He}{ii}\lambda1640) = 0.3$ for Jil2 in Paper I is not compatible with WNh stars.

Therefore, the initial AGN interpretation proposed in Papers I and II for the strong \ion{He}{ii} remains the most viable option. This is further supported by the high \ion{He}{ii} luminosity of $\log(L_{\ion{He}{ii}}/[\mathrm{erg\,s}^{-1}])=42.24\pm0.05$, whereas young massive galaxies are usually limited to \ion{He}{ii} luminosities of $\log(L_{\ion{He}{ii}}/[\mathrm{erg\,s}^{-1}])<41.5$ \citep[e.g.][]{Cassata:2013}. While \ion{He}{ii} nebulae of AGN have been mainly characterised around high-redshift radio galaxies \citep[e.g.][]{Villar-Martin:2007}, a couple of obscured radio-quiet AGN with associated \ion{He}{ii} emission have been identified and studied at $z>2$ \citep[e.g.][]{Dey:2005, Prescott:2009, Cassata:2013, ArrigoniBattaia:2018, denBrok:2020}. The \ion{He}{ii} luminosity in those cases are typically $\log(L_{\ion{He}{ii}}/[\mathrm{erg\,s}^{-1}])\gtrsim42.0$ with a reported line FWHM ranging from 400 to 2200\,$\mathrm{km\,s}^{-1}$.  The integrated \ion{He}{ii} luminosity of Jil2 is clearly in the regime of those  obscured AGN at intermediate and high redshifts, though the line width is at the narrow end. This may be related to the rather compact \ion{He}{ii} emission as potential faint extended emission seen in deeper observations of other AGN was not recovered by us in Paper I. The stellar masses of such radio-quiet AGN host galaxies are often inaccessible or not inferred, but \citet{Cassata:2013} reported a mean stellar mass of $\log(M_\star/\mathrm{M}_\sun)=10.7$ with a significant dispersion among their three studied objects. The stellar mass of Jil2 is lower but still consistent with those cases given the significant uncertainties and scatter in the host masses. While luminous AGN such as LBQS~0302$-$0019 are statistically not necessarily associated with mergers or interacting galaxies at intermediate redshifts \citep[e.g.][]{Schawinski:2012, Mechtley:2016, Villforth:2017, Marian:2019}, it is still a potential channel for coordinated galaxy and black hole growth as AGN luminosity, star formation, and gas content seem to be linked \citep[e.g.][]{Rosario:2013, Husemann:2017, Zhuang:2021}. Given their proximity, Jil1 and Jil2 are interacting galaxies, which presumably triggered the starburst in Jil2. The ongoing or very recent AGN phase as probed by the strong \ion{He}{ii} is likely linked to this starburst phase considering that the typical time delay between the starburst and the AGN phase is of the order of a hundred megayears \citep[e.g.][]{Davies:2007,Wild:2010, Hopkins:2012a}. While the AGN excitation for the bright \ion{He}{ii} line in Jil2 is clearly the favoured option, the stellar population might still contribute a minor fraction to the \ion{He}{ii} ionisation.

Despite the proximity of Jil2 to the luminous QSO LBQS~0302$-$0019, we have already argued against a simple QSO fluorescence scenario for the production of \ion{He}{ii}\,$\lambda$1640 in Paper I. The incident ionising flux of the QSO is only sufficient with the observed \ion{He}{ii} line luminosity for extreme assumptions (e.g. all photons ionising only helium and not hydrogen), which should not be valid. Our conclusion was that there must be an intrinsic source in Jil2 that provides the necessary photons requiring an AGN bolometric luminosity as low as $L_\mathrm{AGN}\sim10^{45}\,\mathrm{erg\,s}^{-1}$.  The young age of Jil2, a few tens of megayears, requires a rapid evolution of the host galaxies, which also implicitly demands significant black hole growth during this period in the co-evolution picture \citep[e.g.][]{Kormendy:2013}. 

\section{Summary and conclusion}
In this paper we have presented new NIR \textit{HST} images of the dual AGN system Jil+LBQS~0302$-$0019 taken with WFC3/IR in the F125W filter ($J$ band). We identified four distinct galaxies close to LBQS~0302$-$0019, at least two of which are physically associated with the QSO. Combining the previous VLT/HAWK-I $K_s$ photometry presented in Paper II with the \textit{HST} $J$ band allowed us to construct the corresponding integrated rest-frame $U-V$ colours for each component. When we compared these to stellar population models, we inferred light-weighted stellar population ages of $<$300\,Myr, suggesting that the companions are rapidly evolving in a group environment. Jil1 is the oldest and most massive galaxy of the companions, with $t_\star=252_{-109}^{+222}\,\mathrm{Myr}$ and $\log(M_\star/[M_\sun])=11.2^{+0.3}_{-0.1}$. This galaxy corresponds to the previously identified companion in ground-based VLT/HAWK-I\ changed{observation}, as initially identified in Paper II. 

More importantly, we were able to robustly associate the strong nebular \ion{He}{ii} emission, initially discovered with MUSE in Paper I, with the companion Jil2 rather than the more massive companion Jil1. Although we inferred a young age of only $\sim$40\,Myr for Jil2 based on the $U-V$ colour, a stellar origin for the \ion{He}{ii} emission could be ruled out given that the \ion{He}{ii} line luminosity, width, and UV line ratios reported in Paper~I cannot be reproduced with current stellar evolution models. As already suggested in Papers I and II, an AGN in Jil2 remains the only viable explanation for the \ion{He}{ii} emission, but a minor stellar contribution is possible.

The inferred age of Jil2 tightly constrains the timing of the intense starburst with a maximum delay of the AGN phase to $t_\star=19_{-14}^{+74}\,\mathrm{Myr}$. This allows us to speculate about the origin of the exceptionally large ($13.2$\,Mpc) \ion{He}{ii} proximity zone around LBQS~0302$-$0019 that has been used to infer a lower limit to the duration of its current radiatively efficient accretion episode of $>11$\,Myr \citep{Worseck:2021}. However, such modelling depends on the UV background radiation field and assumes that only the luminous QSO creates the likely irregular bubble of fully ionised helium, of which we observe a single line of sight to the QSO. The obscured AGN in Jil2 may be acting as an additional source of \ion{He}{ii}-ionising photons, potentially in various directions due to the galaxy motion in the last few megayears, creating the exceptionally large \ion{He}{ii} proximity zone around LBQS~0302$-$0019. Consequently, the duration of the QSO episode of LBQS~0302$-$0019 may have been overestimated. A test of this hypothesis demands VLT/MUSE and deep imaging observations for a statistical sample of QSOs with observable \ion{He}{ii} proximity zones to uncover a potential link between the specific environment properties and the QSO proximity zone sizes.

\begin{acknowledgements}
     The authors greatly appreciate the constructive and valuable feedback from the referee which substantially improved the quality and clarity of the manuscript.
      BH acknowledges financial support by DLR and DFG under grant number 50OR1911 and GE625/17-1, respectively. AACS is Öpik Research Fellow at Armagh Observatory \& Planetarium (AOP). AOP is core funded by the Northern Ireland Government through the Department for Communities. We thank Helge Todt for helpful discussions about Wolf-Rayet stars and galaxies. Based on observations made with the NASA/ESA Hubble Space Telescope, obtained at the Space Telescope Science Institute, which is operated by the Association of Universities for Research in Astronomy, Inc., under NASA contract NAS5-26555. These observations are associated with programme \#15480. 
      Based on observations collected at the European Southern Observatory under ESO programme(s) 60.A-9471(A) and 094.A-0767(A). 
\end{acknowledgements}

\bibliographystyle{aa}
\bibliography{references}

\end{document}

%% file: Table1.tex
\begin{tabular}{cccccccccc}\hline\hline
 & $n$\tablefootmark{a} & $r_\mathrm{e}$\tablefootmark{b} & $b/a$\tablefootmark{c} & $m_\mathrm{F125W}$ (AB)\tablefootmark{d} & $m_\mathrm{Ks}$ (AB)\tablefootmark{e} & $m_\mathrm{F125W}-m_\mathrm{Ks}$ & $t_\mathrm{\star}$\tablefootmark{f} & $\log(M_\star/L_V)$\tablefootmark{g} & $\log(M_\star)$\tablefootmark{h} \\
 & & &  & [mag] & [mag] & [mag] & [Myr] &  & [M$_\sun$]\\\hline
Jil1 & $3.8\pm0.5$ & $0\farcs75\pm0\farcs12$& $0.60\pm0.05$ & $23.70\pm0.27$ & $22.22\pm0.03$ & $1.48\pm0.30$ & $252^{+222}_{-109}$ & $-0.8^{+0.3}_{-0.1}$ & $11.2^{+0.3}_{-0.1}$\\
Jil2 & $1.3\pm0.5$ & $0\farcs76\pm0\farcs65$& $0.26\pm0.07$ & $24.74\pm0.53$ & $24.75\pm0.23$ & $-0.01\pm0.75$ & $19^{+74}_{-14}$ & $-1.5^{+0.9}_{-0.5}$ & $9.4^{+0.9}_{-0.4}$\\
Jil3 & $1.6\pm0.9$ & $0\farcs13\pm0\farcs11$& $0.54\pm0.08$ & $25.69\pm0.25$ & $>25.78$ & $<0.40$ & $<59$ & $<-0.7$ & $<9.9$\\
Jil4 & $0.7\pm0.1$ & $0\farcs26\pm0\farcs01$& $0.60\pm0.04$ & $24.81\pm0.11$ & $24.02\pm0.09$ & $0.80\pm0.19$ & $98^{+41}_{-27}$ & $-1.0^{+0.4}_{-0.1}$ & $10.2^{+0.4}_{-0.1}$\\
\noalign{\smallskip}\hline
\end{tabular}
\tablefoot{
\tablefoottext{a}{best-fit Sersic index estimated from the HST image}
\tablefoottext{b}{best-fit effetive radius estimated from the HST image}
\tablefoottext{c}{best-fit axis-ratio estimated from the HST image}
\tablefoottext{d}{best-fit integrated brightness in the HST image}
\tablefoottext{e}{best-fit integrated brightness in the HAWKI-I  image assuming morphological parameters inferred from the HST image}
\tablefoottext{f}{best-fit luminosity-weighted age of single-stellar population model from the observed colors without dust attenuation}
\tablefoottext{g}{best-fit mass-to-light ratio of the best-fit single-stellar population model}
\tablefoottext{h}{best-fit integrated stellar mass given the luminosity and age of the single-stellar population}
}

%% file: HST_LBQS0302.bbl
\begin{thebibliography}{80}
\expandafter\ifx\csname natexlab\endcsname\relax\def\natexlab#1{#1}\fi

\bibitem[{{Arrigoni Battaia} {et~al.}(2019){Arrigoni Battaia}, {Hennawi},
  {Prochaska}, {O{\~n}orbe}, {Farina}, {Cantalupo}, \&
  {Lusso}}]{ArrigoniBattaia:2019}
{Arrigoni Battaia}, F., {Hennawi}, J.~F., {Prochaska}, J.~X., {et~al.} 2019,
  \mnras, 482, 3162

\bibitem[{{Arrigoni Battaia} {et~al.}(2018){Arrigoni Battaia}, {Prochaska},
  {Hennawi}, {Obreja}, {Buck}, {Cantalupo}, {Dutton}, \&
  {Macci{\`o}}}]{ArrigoniBattaia:2018}
{Arrigoni Battaia}, F., {Prochaska}, J.~X., {Hennawi}, J.~F., {et~al.} 2018,
  \mnras, 473, 3907

\bibitem[{{Avila} {et~al.}(2015){Avila}, {Hack}, {Cara}, {Borncamp}, {Mack},
  {Smith}, \& {Ubeda}}]{Avila:2015}
{Avila}, R.~J., {Hack}, W., {Cara}, M., {et~al.} 2015, in Astronomical Society
  of the Pacific Conference Series, Vol. 495, Astronomical Data Analysis
  Software an Systems XXIV (ADASS XXIV), ed. A.~R. {Taylor} \& E.~{Rosolowsky},
  281

\bibitem[{{Bianchi} {et~al.}(2008){Bianchi}, {Chiaberge}, {Piconcelli},
  {Guainazzi}, \& {Matt}}]{Bianchi:2008}
{Bianchi}, S., {Chiaberge}, M., {Piconcelli}, E., {Guainazzi}, M., \& {Matt},
  G. 2008, \mnras, 386, 105

\bibitem[{{Borisova} {et~al.}(2016){Borisova}, {Cantalupo}, {Lilly}, {Marino},
  {Gallego}, {Bacon}, {Blaizot}, {Bouch{\'e}}, {Brinchmann}, {Carollo},
  {Caruana}, {Finley}, {Herenz}, {Richard}, {Schaye}, {Straka}, {Turner},
  {Urrutia}, {Verhamme}, \& {Wisotzki}}]{Borisova:2016}
{Borisova}, E., {Cantalupo}, S., {Lilly}, S.~J., {et~al.} 2016, \apj, 831, 39

\bibitem[{{Brinchmann} {et~al.}(2008){Brinchmann}, {Kunth}, \&
  {Durret}}]{Brinchmann:2008}
{Brinchmann}, J., {Kunth}, D., \& {Durret}, F. 2008, \aap, 485, 657

\bibitem[{{Bruzual} \& {Charlot}(2003)}]{Bruzual:2003}
{Bruzual}, G. \& {Charlot}, S. 2003, \mnras, 344, 1000

\bibitem[{{Cassata} {et~al.}(2013){Cassata}, {Le F{\`e}vre}, {Charlot},
  {Contini}, {Cucciati}, {Garilli}, {Zamorani}, {Adami}, {Bardelli}, {Le Brun},
  {Lemaux}, {Maccagni}, {Pollo}, {Pozzetti}, {Tresse}, {Vergani}, {Zanichelli},
  \& {Zucca}}]{Cassata:2013}
{Cassata}, P., {Le F{\`e}vre}, O., {Charlot}, S., {et~al.} 2013, \aap, 556, A68

\bibitem[{{Crowther}(2007)}]{Crowther:2007}
{Crowther}, P.~A. 2007, \araa, 45, 177

\bibitem[{{Davies} {et~al.}(2007){Davies}, {M{\"u}ller S{\'a}nchez}, {Genzel},
  {Tacconi}, {Hicks}, {Friedrich}, \& {Sternberg}}]{Davies:2007}
{Davies}, R.~I., {M{\"u}ller S{\'a}nchez}, F., {Genzel}, R., {et~al.} 2007,
  \apj, 671, 1388

\bibitem[{{De Rosa} {et~al.}(2019){De Rosa}, {Vignali}, {Bogdanovi{\'c}},
  {Capelo}, {Charisi}, {Dotti}, {Husemann}, {Lusso}, {Mayer}, {Paragi},
  {Runnoe}, {Sesana}, {Steinborn}, {Bianchi}, {Colpi}, {del Valle}, {Frey},
  {Gab{\'a}nyi}, {Giustini}, {Guainazzi}, {Haiman}, {Herrera Ruiz},
  {Herrero-Illana}, {Iwasawa}, {Komossa}, {Lena}, {Loiseau}, {Perez-Torres},
  {Piconcelli}, \& {Volonteri}}]{deRosa:2019}
{De Rosa}, A., {Vignali}, C., {Bogdanovi{\'c}}, T., {et~al.} 2019, \nar, 86,
  101525

\bibitem[{{Decarli} {et~al.}(2017){Decarli}, {Walter}, {Venemans},
  {Ba{\~n}ados}, {Bertoldi}, {Carilli}, {Fan}, {Farina}, {Mazzucchelli},
  {Riechers}, {Rix}, {Strauss}, {Wang}, \& {Yang}}]{Decarli:2017}
{Decarli}, R., {Walter}, F., {Venemans}, B.~P., {et~al.} 2017, \nat, 545, 457

\bibitem[{{den Brok} {et~al.}(2020){den Brok}, {Cantalupo}, {Mackenzie},
  {Marino}, {Pezzulli}, {Matthee}, {Johnson}, {Krumpe}, {Urrutia}, \&
  {Kollatschny}}]{denBrok:2020}
{den Brok}, J.~S., {Cantalupo}, S., {Mackenzie}, R., {et~al.} 2020, \mnras,
  495, 1874

\bibitem[{{Dey} {et~al.}(2005){Dey}, {Bian}, {Soifer}, {Brand}, {Brown},
  {Chaffee}, {Le Floc'h}, {Hill}, {Houck}, {Jannuzi}, {Rieke}, {Weedman},
  {Brodwin}, \& {Eisenhardt}}]{Dey:2005}
{Dey}, A., {Bian}, C., {Soifer}, B.~T., {et~al.} 2005, \apj, 629, 654

\bibitem[{{Djorgovski} {et~al.}(2007){Djorgovski}, {Courbin}, {Meylan},
  {Sluse}, {Thompson}, {Mahabal}, \& {Glikman}}]{Djorgovski:2007}
{Djorgovski}, S.~G., {Courbin}, F., {Meylan}, G., {et~al.} 2007, \apjl, 662, L1

\bibitem[{{Drake} {et~al.}(2019){Drake}, {Farina}, {Neeleman}, {Walter},
  {Venemans}, {Banados}, {Mazzucchelli}, \& {Decarli}}]{Drake:2019}
{Drake}, A.~B., {Farina}, E.~P., {Neeleman}, M., {et~al.} 2019, \apj, 881, 131

\bibitem[{{Farina} {et~al.}(2019){Farina}, {Arrigoni-Battaia}, {Costa},
  {Walter}, {Hennawi}, {Drake}, {Decarli}, {Gutcke}, {Mazzucchelli},
  {Neeleman}, {Georgiev}, {Eilers}, {Davies}, {Ba{\~n}ados}, {Fan}, {Onoue},
  {Schindler}, {Venemans}, {Wang}, {Yang}, {Rabien}, \& {Busoni}}]{Farina:2019}
{Farina}, E.~P., {Arrigoni-Battaia}, F., {Costa}, T., {et~al.} 2019, \apj, 887,
  196

\bibitem[{{Findlay} {et~al.}(2018){Findlay}, {Prochaska}, {Hennawi},
  {Fumagalli}, {Myers}, {Bartle}, {Chehade}, {DiPompeo}, {Shanks}, {Lau}, \&
  {Rubin}}]{Findlay:2018}
{Findlay}, J.~R., {Prochaska}, J.~X., {Hennawi}, J.~F., {et~al.} 2018, \apjs,
  236, 44

\bibitem[{{Frey} \& {Gab{\'a}nyi}(2018)}]{Frey:2018}
{Frey}, S. \& {Gab{\'a}nyi}, K.~{\'E}. 2018, Research Notes of the AAS, 2, 49

\bibitem[{{Fu} {et~al.}(2012){Fu}, {Yan}, {Myers}, {Stockton}, {Djorgovski},
  {Aldering}, \& {Rich}}]{Fu:2012}
{Fu}, H., {Yan}, L., {Myers}, A.~D., {et~al.} 2012, \apj, 745, 67

\bibitem[{{Georgakakis} {et~al.}(2017){Georgakakis}, {Aird}, {Schulze},
  {Dwelly}, {Salvato}, {Nandra}, {Merloni}, \& {Schneider}}]{Georgakakis:2017}
{Georgakakis}, A., {Aird}, J., {Schulze}, A., {et~al.} 2017, \mnras, 471, 1976

\bibitem[{{Gr{\"a}fener} \& {Vink}(2015)}]{Graefener:2015}
{Gr{\"a}fener}, G. \& {Vink}, J.~S. 2015, \aap, 578, L2

\bibitem[{{Gross} {et~al.}(2019){Gross}, {Fu}, {Myers}, {Wrobel}, \&
  {Djorgovski}}]{Gross:2019}
{Gross}, A.~C., {Fu}, H., {Myers}, A.~D., {Wrobel}, J.~M., \& {Djorgovski},
  S.~G. 2019, \apj, 883, 50

\bibitem[{{Guainazzi} {et~al.}(2005){Guainazzi}, {Piconcelli},
  {Jim{\'e}nez-Bail{\'o}n}, \& {Matt}}]{Guainazzi:2005}
{Guainazzi}, M., {Piconcelli}, E., {Jim{\'e}nez-Bail{\'o}n}, E., \& {Matt}, G.
  2005, \aap, 429, L9

\bibitem[{{Hatch} {et~al.}(2014){Hatch}, {Wylezalek}, {Kurk}, {Stern}, {De
  Breuck}, {Jarvis}, {Galametz}, {Gonzalez}, {Hartley}, {Mortlock}, {Seymour},
  \& {Stevens}}]{Hatch:2014}
{Hatch}, N.~A., {Wylezalek}, D., {Kurk}, J.~D., {et~al.} 2014, \mnras, 445, 280

\bibitem[{{Heckman} {et~al.}(1991){Heckman}, {Miley}, {Lehnert}, \& {van
  Breugel}}]{Heckman:1991b}
{Heckman}, T.~M., {Miley}, G.~K., {Lehnert}, M.~D., \& {van Breugel}, W. 1991,
  \apj, 370, 78

\bibitem[{{Hennawi} {et~al.}(2015){Hennawi}, {Prochaska}, {Cantalupo}, \&
  {Arrigoni-Battaia}}]{Hennawi:2015}
{Hennawi}, J.~F., {Prochaska}, J.~X., {Cantalupo}, S., \& {Arrigoni-Battaia},
  F. 2015, Science, 348, 779

\bibitem[{{Hopkins}(2012)}]{Hopkins:2012a}
{Hopkins}, P.~F. 2012, \mnras, 420, L8

\bibitem[{{Hopkins} {et~al.}(2008){Hopkins}, {Hernquist}, {Cox}, \& {Kere{\v
  s}}}]{Hopkins:2008a}
{Hopkins}, P.~F., {Hernquist}, L., {Cox}, T.~J., \& {Kere{\v s}}, D. 2008,
  \apjs, 175, 356

\bibitem[{{Hu} {et~al.}(1995){Hu}, {Kim}, {Cowie}, {Songaila}, \&
  {Rauch}}]{Hu:1995}
{Hu}, E.~M., {Kim}, T.-S., {Cowie}, L.~L., {Songaila}, A., \& {Rauch}, M. 1995,
  \aj, 110, 1526

\bibitem[{{Husemann} {et~al.}(2018{\natexlab{a}}){Husemann}, {Bielby},
  {Jahnke}, {Arrigoni-Battaia}, {Worseck}, {Shanks}, {Wardlow}, \&
  {Scholtz}}]{Husemann:2018c}
{Husemann}, B., {Bielby}, R., {Jahnke}, K., {et~al.} 2018{\natexlab{a}}, \aap,
  614, L2

\bibitem[{{Husemann} {et~al.}(2017){Husemann}, {Davis}, {Jahnke},
  {Dannerbauer}, {Urrutia}, \& {Hodge}}]{Husemann:2017}
{Husemann}, B., {Davis}, T.~A., {Jahnke}, K., {et~al.} 2017, \mnras, 470, 1570

\bibitem[{{Husemann} {et~al.}(2020){Husemann}, {Heidt}, {De Rosa}, {Vignali},
  {Bianchi}, {Bogdanovi{\'c}}, {Komossa}, \& {Paragi}}]{Husemann:2020}
{Husemann}, B., {Heidt}, J., {De Rosa}, A., {et~al.} 2020, \aap, 639, A117

\bibitem[{{Husemann} {et~al.}(2018{\natexlab{b}}){Husemann}, {Worseck},
  {Arrigoni Battaia}, \& {Shanks}}]{Husemann:2018}
{Husemann}, B., {Worseck}, G., {Arrigoni Battaia}, F., \& {Shanks}, T.
  2018{\natexlab{b}}, \aap, 610, L7

\bibitem[{{Jakobsen} {et~al.}(1994){Jakobsen}, {Boksenberg}, {Deharveng},
  {Greenfield}, {Jedrzejewski}, \& {Paresce}}]{Jakobsen:1994}
{Jakobsen}, P., {Boksenberg}, A., {Deharveng}, J.~M., {et~al.} 1994, \nat, 370,
  35

\bibitem[{{Jakobsen} {et~al.}(2003){Jakobsen}, {Jansen}, {Wagner}, \&
  {Reimers}}]{Jakobsen:2003}
{Jakobsen}, P., {Jansen}, R.~A., {Wagner}, S., \& {Reimers}, D. 2003, \aap,
  397, 891

\bibitem[{{Kehrig} {et~al.}(2018){Kehrig}, {V{\'\i}lchez}, {Guerrero},
  {Iglesias-P{\'a}ramo}, {Hunt}, {Duarte-Puertas}, \&
  {Ramos-Larios}}]{Kehrig:2018}
{Kehrig}, C., {V{\'\i}lchez}, J.~M., {Guerrero}, M.~A., {et~al.} 2018, \mnras,
  480, 1081

\bibitem[{{Kehrig} {et~al.}(2015){Kehrig}, {V{\'\i}lchez}, {P{\'e}rez-Montero},
  {Iglesias-P{\'a}ramo}, {Brinchmann}, {Kunth}, {Durret}, \&
  {Bayo}}]{Kehrig:2015}
{Kehrig}, C., {V{\'\i}lchez}, J.~M., {P{\'e}rez-Montero}, E., {et~al.} 2015,
  \apjl, 801, L28

\bibitem[{{Komossa} {et~al.}(2003){Komossa}, {Burwitz}, {Hasinger}, {Predehl},
  {Kaastra}, \& {Ikebe}}]{Komossa:2003}
{Komossa}, S., {Burwitz}, V., {Hasinger}, G., {et~al.} 2003, \apjl, 582, L15

\bibitem[{{Kormendy} \& {Ho}(2013)}]{Kormendy:2013}
{Kormendy}, J. \& {Ho}, L.~C. 2013, \araa, 51, 511

\bibitem[{{Koss} {et~al.}(2011){Koss}, {Mushotzky}, {Treister}, {Veilleux},
  {Vasudevan}, {Miller}, {Sanders}, {Schawinski}, \& {Trippe}}]{Koss:2011}
{Koss}, M., {Mushotzky}, R., {Treister}, E., {et~al.} 2011, \apjl, 735, L42

\bibitem[{{Liang} {et~al.}(2020){Liang}, {Li}, {Li}, {Yan}, {Mo}, {Zhang},
  {Machuca}, \& {Roman-Lopes}}]{Liang:2020}
{Liang}, F.-H., {Li}, C., {Li}, N., {et~al.} 2020, \apj, 896, 121

\bibitem[{{Marconi} {et~al.}(2004){Marconi}, {Risaliti}, {Gilli}, {Hunt},
  {Maiolino}, \& {Salvati}}]{Marconi:2004}
{Marconi}, A., {Risaliti}, G., {Gilli}, R., {et~al.} 2004, \mnras, 351, 169

\bibitem[{{Marian} {et~al.}(2019){Marian}, {Jahnke}, {Mechtley}, {Cohen},
  {Husemann}, {Jones}, {Koekemoer}, {Schulze}, {van der Wel}, {Villforth}, \&
  {Windhorst}}]{Marian:2019}
{Marian}, V., {Jahnke}, K., {Mechtley}, M., {et~al.} 2019, \apj, 882, 141

\bibitem[{{Mechtley} {et~al.}(2016){Mechtley}, {Jahnke}, {Windhorst}, {Andrae},
  {Cisternas}, {Cohen}, {Hewlett}, {Koekemoer}, {Schramm}, {Schulze},
  {Silverman}, {Villforth}, {van der Wel}, \& {Wisotzki}}]{Mechtley:2016}
{Mechtley}, M., {Jahnke}, K., {Windhorst}, R.~A., {et~al.} 2016, \apj, 830, 156

\bibitem[{{Miralles-Caballero} {et~al.}(2016){Miralles-Caballero}, {D{\'\i}az},
  {L{\'o}pez-S{\'a}nchez}, {Rosales-Ortega}, {Monreal-Ibero},
  {P{\'e}rez-Montero}, {Kehrig}, {Garc{\'\i}a-Benito}, {S{\'a}nchez},
  {Walcher}, {Galbany}, {Iglesias-P{\'a}ramo}, {V{\'\i}lchez}, {Gonz{\'a}lez
  Delgado}, {van de Ven}, {Barrera-Ballesteros}, {Lyubenova}, {Meidt},
  {Falcon-Barroso}, {Mast}, {Mendoza}, \& {Califa
  Collaboration}}]{Miralles-Caballero:2016}
{Miralles-Caballero}, D., {D{\'\i}az}, A.~I., {L{\'o}pez-S{\'a}nchez},
  {\'A}.~R., {et~al.} 2016, \aap, 592, A105

\bibitem[{{Nanayakkara} {et~al.}(2019){Nanayakkara}, {Brinchmann}, {Boogaard},
  {Bouwens}, {Cantalupo}, {Feltre}, {Kollatschny}, {Marino}, {Maseda},
  {Matthee}, {Paalvast}, {Richard}, \& {Verhamme}}]{Nanyakkara:2019}
{Nanayakkara}, T., {Brinchmann}, J., {Boogaard}, L., {et~al.} 2019, \aap, 624,
  A89

\bibitem[{{O'Sullivan} {et~al.}(2020){O'Sullivan}, {Martin}, {Matuszewski},
  {Hoadley}, {Hamden}, {Neill}, {Lin}, \& {Parihar}}]{OSullivan:2020}
{O'Sullivan}, D.~B., {Martin}, C., {Matuszewski}, M., {et~al.} 2020, \apj, 894,
  3

\bibitem[{{Peng} {et~al.}(2010){Peng}, {Ho}, {Impey}, \& {Rix}}]{Peng:2010}
{Peng}, C.~Y., {Ho}, L.~C., {Impey}, C.~D., \& {Rix}, H. 2010, \aj, 139, 2097

\bibitem[{{Peng} {et~al.}(2002){Peng}, {Ho}, {Impey}, \& {Rix}}]{Peng:2002}
{Peng}, C.~Y., {Ho}, L.~C., {Impey}, C.~D., \& {Rix}, H.-W. 2002, \aj, 124, 266

\bibitem[{{Piconcelli} {et~al.}(2010){Piconcelli}, {Vignali}, {Bianchi},
  {Mathur}, {Fiore}, {Guainazzi}, {Lanzuisi}, {Maiolino}, \&
  {Nicastro}}]{Piconcelli:2010}
{Piconcelli}, E., {Vignali}, C., {Bianchi}, S., {et~al.} 2010, \apjl, 722, L147

\bibitem[{{Prescott} {et~al.}(2009){Prescott}, {Dey}, \&
  {Jannuzi}}]{Prescott:2009}
{Prescott}, M. K.~M., {Dey}, A., \& {Jannuzi}, B.~T. 2009, \apj, 702, 554

\bibitem[{{Rosario} {et~al.}(2013){Rosario}, {Trakhtenbrot}, {Lutz}, {Netzer},
  {Trump}, {Silverman}, {Schramm}, {Lusso}, {Berta}, {Bongiorno}, {Brusa},
  {F{\"o}rster-Schreiber}, {Genzel}, {Lilly}, {Magnelli}, {Mainieri},
  {Maiolino}, {Merloni}, {Mignoli}, {Nordon}, {Popesso}, {Salvato}, {Santini},
  {Tacconi}, \& {Zamorani}}]{Rosario:2013}
{Rosario}, D.~J., {Trakhtenbrot}, B., {Lutz}, D., {et~al.} 2013, \aap, 560, A72

\bibitem[{{Sander} \& {Vink}(2020)}]{Sander:2020}
{Sander}, A. A.~C. \& {Vink}, J.~S. 2020, \mnras, 499, 873

\bibitem[{{Sanders} {et~al.}(1988){Sanders}, {Soifer}, {Elias}, {Madore},
  {Matthews}, {Neugebauer}, \& {Scoville}}]{Sanders:1988a}
{Sanders}, D.~B., {Soifer}, B.~T., {Elias}, J.~H., {et~al.} 1988, \apj, 325, 74

\bibitem[{{Saxena} {et~al.}(2020){Saxena}, {Pentericci}, {Mirabelli},
  {Schaerer}, {Schneider}, {Cullen}, {Amorin}, {Bolzonella}, {Bongiorno},
  {Carnall}, {Castellano}, {Cucciati}, {Fontana}, {Fynbo}, {Garilli},
  {Gargiulo}, {Guaita}, {Hathi}, {Hutchison}, {Koekemoer}, {Marchi}, {McLeod},
  {McLure}, {Papovich}, {Pozzetti}, {Talia}, \& {Zamorani}}]{Saxena:2020}
{Saxena}, A., {Pentericci}, L., {Mirabelli}, M., {et~al.} 2020, \aap, 636, A47

\bibitem[{{Schaerer} \& {Vacca}(1998)}]{Schaerer:1998}
{Schaerer}, D. \& {Vacca}, W.~D. 1998, \apj, 497, 618

\bibitem[{{Schawinski} {et~al.}(2012){Schawinski}, {Simmons}, {Urry},
  {Treister}, \& {Glikman}}]{Schawinski:2012}
{Schawinski}, K., {Simmons}, B.~D., {Urry}, C.~M., {Treister}, E., \&
  {Glikman}, E. 2012, \mnras, 425, L61

\bibitem[{{Schmidt} {et~al.}(2017){Schmidt}, {Worseck}, {Hennawi}, {Prochaska},
  \& {Crighton}}]{Schmidt:2017}
{Schmidt}, T.~M., {Worseck}, G., {Hennawi}, J.~F., {Prochaska}, J.~X., \&
  {Crighton}, N.~H.~M. 2017, \apj, 847, 81

\bibitem[{{Schulze} {et~al.}(2015){Schulze}, {Bongiorno}, {Gavignaud},
  {Schramm}, {Silverman}, {Merloni}, {Zamorani}, {Hirschmann}, {Mainieri},
  {Wisotzki}, {Shankar}, {Fiore}, {Koekemoer}, \& {Temporin}}]{Schulze:2015}
{Schulze}, A., {Bongiorno}, A., {Gavignaud}, I., {et~al.} 2015, \mnras, 447,
  2085

\bibitem[{{Senchyna} {et~al.}(2020){Senchyna}, {Stark}, {Mirocha}, {Reines},
  {Charlot}, {Jones}, \& {Mulchaey}}]{Senchyna:2020}
{Senchyna}, P., {Stark}, D.~P., {Mirocha}, J., {et~al.} 2020, \mnras, 494, 941

\bibitem[{{Silverman} {et~al.}(2020){Silverman}, {Tang}, {Lee}, {Hartwig},
  {Goulding}, {Strauss}, {Schramm}, {Ding}, {Riffel}, {Fujimoto}, {Hikage},
  {Imanishi}, {Iwasawa}, {Jahnke}, {Kayo}, {Kashikawa}, {Kawaguchi}, {Kohno},
  {Luo}, {Matsuoka}, {Matsuda}, {Nagao}, {Oguri}, {Ono}, {Onoue}, {Ouchi},
  {Shimasaku}, {Suh}, {Suzuki}, {Taniguchi}, {Toba}, {Ueda}, \&
  {Yasuda}}]{Silverman:2020}
{Silverman}, J.~D., {Tang}, S., {Lee}, K.-G., {et~al.} 2020, \apj, 899, 154

\bibitem[{{Soltan}(1982)}]{Soltan:1982}
{Soltan}, A. 1982, \mnras, 200, 115

\bibitem[{{Steidel} {et~al.}(2003){Steidel}, {Adelberger}, {Shapley},
  {Pettini}, {Dickinson}, \& {Giavalisco}}]{Steidel:2003}
{Steidel}, C.~C., {Adelberger}, K.~L., {Shapley}, A.~E., {et~al.} 2003, \apj,
  592, 728

\bibitem[{{Stott} {et~al.}(2020){Stott}, {Bielby}, {Cullen}, {Burchett},
  {Tejos}, {Fumagalli}, {Crain}, {Morris}, {Amos}, {Bower}, \&
  {Prochaska}}]{Stott:2020}
{Stott}, J.~P., {Bielby}, R.~M., {Cullen}, F., {et~al.} 2020, \mnras, 497, 3083

\bibitem[{{Syphers} \& {Shull}(2014)}]{Syphers:2014}
{Syphers}, D. \& {Shull}, J.~M. 2014, \apj, 784, 42

\bibitem[{{Trakhtenbrot} {et~al.}(2017){Trakhtenbrot}, {Lira}, {Netzer},
  {Cicone}, {Maiolino}, \& {Shemmer}}]{Trakhtenbrot:2017}
{Trakhtenbrot}, B., {Lira}, P., {Netzer}, H., {et~al.} 2017, \apj, 836, 8

\bibitem[{{Tummuangpak} {et~al.}(2014){Tummuangpak}, {Bielby}, {Shanks},
  {Theuns}, {Crighton}, {Francke}, \& {Infante}}]{Tummuangpak:2014}
{Tummuangpak}, P., {Bielby}, R.~M., {Shanks}, T., {et~al.} 2014, \mnras, 442,
  2094

\bibitem[{{Venemans} {et~al.}(2007){Venemans}, {R{\"o}ttgering}, {Miley}, {van
  Breugel}, {de Breuck}, {Kurk}, {Pentericci}, {Stanford}, {Overzier}, {Croft},
  \& {Ford}}]{Venemans:2007}
{Venemans}, B.~P., {R{\"o}ttgering}, H.~J.~A., {Miley}, G.~K., {et~al.} 2007,
  \aap, 461, 823

\bibitem[{{Venemans} {et~al.}(2020){Venemans}, {Walter}, {Neeleman}, {Novak},
  {Otter}, {Decarli}, {Ba{\~n}ados}, {Drake}, {Farina}, {Kaasinen},
  {Mazzucchelli}, {Carilli}, {Fan}, {Rix}, \& {Wang}}]{Venemans:2020}
{Venemans}, B.~P., {Walter}, F., {Neeleman}, M., {et~al.} 2020, \apj, 904, 130

\bibitem[{{Villar-Mart{\'{\i}}n} {et~al.}(2007){Villar-Mart{\'{\i}}n},
  {S{\'a}nchez}, {Humphrey}, {Dijkstra}, {di Serego Alighieri}, {De Breuck}, \&
  {Gonz{\'a}lez Delgado}}]{Villar-Martin:2007}
{Villar-Mart{\'{\i}}n}, M., {S{\'a}nchez}, S.~F., {Humphrey}, A., {et~al.}
  2007, \mnras, 378, 416

\bibitem[{{Villforth} {et~al.}(2017){Villforth}, {Hamilton}, {Pawlik},
  {Hewlett}, {Rowlands}, {Herbst}, {Shankar}, {Fontana}, {Hamann}, {Koekemoer},
  {Pforr}, {Trump}, \& {Wuyts}}]{Villforth:2017}
{Villforth}, C., {Hamilton}, T., {Pawlik}, M.~M., {et~al.} 2017, \mnras, 466,
  812

\bibitem[{{Weigel} {et~al.}(2018){Weigel}, {Schawinski}, {Treister},
  {Trakhtenbrot}, \& {Sanders}}]{Weigel:2018}
{Weigel}, A.~K., {Schawinski}, K., {Treister}, E., {Trakhtenbrot}, B., \&
  {Sanders}, D.~B. 2018, \mnras, 476, 2308

\bibitem[{{Wild} {et~al.}(2010){Wild}, {Heckman}, \& {Charlot}}]{Wild:2010}
{Wild}, V., {Heckman}, T., \& {Charlot}, S. 2010, \mnras, 405, 933

\bibitem[{{Wisotzki} {et~al.}(2016){Wisotzki}, {Bacon}, {Blaizot},
  {Brinchmann}, {Herenz}, {Schaye}, {Bouch{\'e}}, {Cantalupo}, {Contini},
  {Carollo}, {Caruana}, {Courbot}, {Emsellem}, {Kamann}, {Kerutt}, {Leclercq},
  {Lilly}, {Patr{\'\i}cio}, {Sandin}, {Steinmetz}, {Straka}, {Urrutia},
  {Verhamme}, {Weilbacher}, \& {Wendt}}]{Wisotzki:2016}
{Wisotzki}, L., {Bacon}, R., {Blaizot}, J., {et~al.} 2016, \aap, 587, A98

\bibitem[{{Wofford} {et~al.}(2014){Wofford}, {Leitherer}, {Chandar}, \&
  {Bouret}}]{Wofford:2014}
{Wofford}, A., {Leitherer}, C., {Chandar}, R., \& {Bouret}, J.-C. 2014, \apj,
  781, 122

\bibitem[{{Woo} {et~al.}(2014){Woo}, {Cho}, {Husemann}, {Komossa}, {Park}, \&
  {Bennert}}]{Woo:2014}
{Woo}, J.-H., {Cho}, H., {Husemann}, B., {et~al.} 2014, \mnras, 437, 32

\bibitem[{{Worseck} {et~al.}(2021){Worseck}, {Khrykin}, {Hennawi}, {Prochaska},
  \& {Farina}}]{Worseck:2021}
{Worseck}, G., {Khrykin}, I.~S., {Hennawi}, J.~F., {Prochaska}, J.~X., \&
  {Farina}, E.~P. 2021, \mnras, 505, 5084

\bibitem[{{Wylezalek} {et~al.}(2013){Wylezalek}, {Galametz}, {Stern}, {Vernet},
  {De Breuck}, {Seymour}, {Brodwin}, {Eisenhardt}, {Gonzalez}, {Hatch},
  {Jarvis}, {Rettura}, {Stanford}, \& {Stevens}}]{Wylezalek:2013}
{Wylezalek}, D., {Galametz}, A., {Stern}, D., {et~al.} 2013, \apj, 769, 79

\bibitem[{{Zhuang} {et~al.}(2021){Zhuang}, {Ho}, \& {Shangguan}}]{Zhuang:2021}
{Zhuang}, M.-Y., {Ho}, L.~C., \& {Shangguan}, J. 2021, \apj, 906, 38

\end{thebibliography}
